\documentclass[sn-mathphys,Numbered]{sn-jnl}

\catcode`\|=12\relax 

\usepackage{amsmath}
\usepackage{amsthm}
\usepackage{amsfonts}
\usepackage{listings}
\usepackage{MnSymbol}
\usepackage{tikz}
\usepackage{hyperref}
\usepackage[utf8x]{inputenc}
\usepackage{upgreek}
\usepackage{xcolor}
\usepackage{colortbl}
\usepackage[inline]{enumitem}
\usepackage[capitalise]{cleveref}
\usepackage{subcaption}
\usepackage{tikz-cd}
\usepackage{tabularx}
\usepackage{flushend}
\usepackage{xspace}
\usepackage{microtype}
\usepackage{scalefnt}
\usepackage{manyfoot} 

\theoremstyle{thmstyleone}%
\newtheorem*{theorem*}{Theorem}
\theoremstyle{thmstyletwo}%

\theoremstyle{thmstylethree}%
\newtheorem*{definition*}{Definition}%

\makeatletter
\global\let\tikz@ensure@dollar@catcode=\relax
\makeatother

\definecolor{keywordcolor}{rgb}{0.7, 0.1, 0.1}   
\definecolor{tacticcolor}{rgb}{0.0, 0.1, 0.3}    
\definecolor{commentcolor}{rgb}{0.4, 0.4, 0.4}   
\definecolor{stringcolor}{rgb}{0.5, 0.3, 0.2}    
\definecolor{symbolcolor}{rgb}{0.1, 0.2, 0.7}    
\definecolor{sortcolor}{rgb}{0.1, 0.5, 0.1}      
\definecolor{attributecolor}{rgb}{0.7, 0.1, 0.1} 
\definecolor{errorcolor}{rgb}{1, 0, 0}           

\DeclareMathOperator {\End}{End}
\DeclareMathOperator {\range}{range}

\DeclareMathOperator {\ad}{ad}

\newcommand{\extlink}[1]{
\href{https://github.com/leanprover-community/mathlib/blob/16728b3064a1751103e1dc2815ed8d00560e0d87/src/#1}{[src]}}

\begin{document}

\title[Engel's theorem in Mathlib]{Engel's theorem in Mathlib}

\author{\fnm{Oliver} \sur{Nash}}\email{o.nash@imperial.ac.uk}

\affil{
  \orgdiv{Department of Mathematics},
  \orgname{Imperial College},
  \orgaddress{\city{London},\country{UK}}}

\abstract{We discuss the theory of Lie algebras in Lean's Mathlib library.
Using nilpotency as the theme, we outline a computer formalisation of Engel's theorem and an
application to root space theory. We emphasise that all arguments work
with coefficients in any commutative ring.}

\keywords{Lean, proof assistant, formal math, Lie algebra, nilpotency, Engel's theorem, root space}

\pacs[MSC Classification]{17B30, 68V15, 68V20}

\maketitle

\section{Introduction}
In previous work \cite{Nash2022}, we reported on the formalisation of Lie algebras in Lean's Mathlib
library \cite{Mathlib}.
Whereas the aim of \cite{Nash2022} was to give a high-level overview of the state of Mathlib's Lie theory,
our goal here is to use a particular aspect of Lie theory (nilpotency) as a means for illustrating
two of Mathlib's most important virtues: generality and unity. For example we shall see that
Mathlib's version of Engel's theorem unifies the various versions of this result appearing in the
informal literature, while demanding weaker hypotheses and providing a stronger result.

This article also presents an opportunity to highlight certain additions to Mathlib's Lie algebra theory
that have been made since \cite{Nash2022}. To be precise, Mathlib's Lie theory has grown from c.6,000
lines to c.7,500 lines since \cite{Nash2022}. The most important additions are the proof of Engel's
theorem and its corollary \eqref{zeroRootCartanIff} characterising Cartan subalgebras.

We briefly summarise some key context and refer the reader to the references for further discussion.
\href{https://leanprover.github.io/}{Lean} is an open-source, dependently-typed functional
programming language and proof assistant. The principal
developer is Leonardo de Moura at Microsoft Research. It uses a type theory similar
to that of Coq for its logical foundation, see e.g., \cite{CarneiroMaster}.

\href{https://leanprover-community.github.io/}{Mathlib} \cite{Mathlib} is an open-source,
massively-collaborative library of formal mathematics written using Lean. As of June 2022,
it contains c.900,000 lines of code, is steadily growing, and contains large quantities
of contemporary undergraduate- and graduate-level mathematics.

Lie algebras are an important class of algebras which arise throughout mathematics and
physics. Well-known to geometers as an abstraction of infinitesimal symmetry, Lie algebras
show up throughout mathematics, occupying prominent roles across the subject, from number theory
to differential geometry. They are also essential for understanding much of 20$^{\rm th}$
Century physics, especially the Standard Model in particle physics.
In addition, the beautiful classification of semisimple Lie algebras and their representations
has made them an object of study in their own right.

As outlined in \cite{Nash2022}, Mathlib contains a substantial body of Lie algebra
theory. The starting point is a pair of typeclasses which permit one to make the statement that a
type $L$ carries the structure of a Lie algebra over\footnote{I.e., with coefficients in.}
a commutative ring $R$. For the benefit of non-specialists we emphasise
that Lie rings are not required to be associative\footnote{Associative Lie rings with a
non-trivial product are rare and rather uninteresting.} and so are not rings in the
sense in which the term is most often used.
The formal definitions in Mathlib are as follows\extlink{algebra/lie/basic.lean\#L54}:
\begin{lstlisting}
class lie_ring (L : Type v)
    extends add_comm_group L, has_bracket L L :=
(add_lie     : ∀ (x y z : L), ⁅x + y,z⁆ = ⁅x,z⁆ + ⁅y,z⁆)
(lie_add     : ∀ (x y z : L), ⁅x,y + z⁆ = ⁅x,y⁆ + ⁅x,z⁆)
(lie_self    : ∀ (x : L), ⁅x,x⁆ = 0)
(leibniz_lie : ∀ (x y z : L), ⁅x,⁅y,z⁆⁆ = ⁅⁅x,y⁆,z⁆ + ⁅y,⁅x,z⁆⁆)

class lie_algebra (R : Type u) (L : Type v)
    [comm_ring R] [lie_ring L] extends module R L :=
(lie_smul : ∀ (t : R) (x y : L), ⁅x,t • y⁆ = t • ⁅x,y⁆)
\end{lstlisting}

We emphasise that no assumptions are made about the coefficients $R$
except that they form a commutative ring. Practictioners of formal mathematics
will be familiar with the value of making definitions as general as reasonably possible.
Of course, if one also seeks to make theorems correspondingly general, then new arguments are
often required. One of the points of this article is that this extra effort
may yield unexpected benefits.

Our case study is Engel's theorem which
turns out to be true for any commutative ring $R$ and to apply to a generalisation of
nilpotency to Lie modules. As we shall see, the generalisation of nilpotency, from algebras
to modules, is especially convenient as it unifies the various versions of Engel's theorem
that appear in the literature. In addition we obtain a slightly more powerful result;
this is useful when it comes to applications and we give an example of this for root
spaces in section \ref{rootSpaceSect}.

\section{Lie modules}
Given a Lie algebra $L$, Mathlib's definition of a Lie module\footnote{A Lie module $M$ is also known as a \emph{representation}
of a Lie algebra $L$ and it is synonymous to say that $M$ is a module {\lq}of{\rq} $L$ or a module
{\lq}over{\rq} $L$.} $M$ of $L$ is\extlink{algebra/lie/basic.lean\#L69}:
\begin{lstlisting}
class lie_ring_module (L : Type v) (M : Type w)
  [lie_ring L] [add_comm_group M] extends has_bracket L M :=
(add_lie     : ∀ (x y : L) (m : M), ⁅x+y,m⁆ = ⁅x,m⁆ + ⁅y,m⁆)
(lie_add     : ∀ (x : L) (m n : M), ⁅x,m+n⁆ = ⁅x,m⁆ + ⁅x,n⁆)
(leibniz_lie : ∀ (x y : L) (m : M),
    ⁅x, ⁅y, m⁆⁆ = ⁅⁅x, y⁆, m⁆ + ⁅y, ⁅x, m⁆⁆)

class lie_module (R : Type u) (L : Type v) (M : Type w)
  [comm_ring R] [lie_ring L] [lie_algebra R L]
  [add_comm_group M] [module R M] [lie_ring_module L M] :=
(smul_lie : ∀ (t : R) (x : L) (m : M), ⁅t • x, m⁆ = t • ⁅x, m⁆)
(lie_smul : ∀ (t : R) (x : L) (m : M), ⁅x, t • m⁆ = t • ⁅x, m⁆)
\end{lstlisting}
Observe that we denote the action of an element $x$ of a Lie algebra $L$ on
an element $m$ of a Lie module $M$ by $[x, m]$. This permits a uniform
notation when regarding a Lie algebra as a module over itself.

The action of $L$ on $M$ induces a natural map:
\begin{align*}
  \phi : L &\to \End(M)\\
  x &\mapsto \phi_x,
\end{align*}
defined by:
\begin{align*}
  \phi_x = [x, m].
\end{align*}
In Mathlib this is\extlink{algebra/lie/of_associative.lean\#L176}:
\begin{lstlisting}
def lie_module.to_endomorphism : L →ₗ⁅R⁆ module.End R M :=
{ to_fun    := λ x,
  { to_fun    := λ m, ⁅x, m⁆,
    -- etc.
    } }
\end{lstlisting}
Any Lie algebra acts on itself\footnote{I.e., is a module over itself.} and in this special case, when $M = L$, the
map $\phi$ is known as the adjoint action, and is denoted\extlink{algebra/lie/of_associative.lean\#L186}:
\begin{align*}
  \ad : L &\to \End(L)\\
  x &\mapsto \ad_x.
\end{align*}

\section{Nilpotency}
Nilpotency is an important concept in algebra, with applications beyond
what one might na\"ively expect from the definitions. In the case of Lie algebras,
the representation theory of nilpotent Lie algebras is both especially
simple and especially important since
semisimple Lie algebras are best understood as representations of certain nilpotent
subalgebras, namely, Cartan subalgebras\footnote{Most of the theory
of semisimple Lie algebras has yet to be formalised. Adding this material to Mathlib
following, say \cite{serre1965}, would be a worthy endeavour.}.

\subsection{Ideal operations and nilpotent Lie modules}
Given a Lie module $M$ of a Lie algebra $L$, recall (\cite{Nash2022} section 3) that for any
ideal $I \subseteq L$ and Lie submodule $N \subseteq M$ we can form a new Lie
submodule\extlink{algebra/lie/ideal_operations.lean\#L48}:
\begin{align*}
  [I, N] \subseteq M.
\end{align*}
It is the smallest Lie submodule of $M$ containing all elements of the form $[x, m]$ for all
$x \in I$ and $m \in N$.

Informally, the lower central series of a Lie module $M$ is the sequence of Lie submodules
of $M$ defined recursively as:
\begin{align*}
  C_0 M     &= M,\\
  C_{k+1} M &= [L, C_k M].
\end{align*}
The corresponding definitions in Mathlib are\extlink{algebra/lie/nilpotent.lean\#L50}:
\begin{lstlisting}
def lcs (k : ℕ) : lie_submodule R L M → lie_submodule R L M :=
(λ N, ⁅( ⊤ : lie_ideal R L), N⁆)^[k]

def lower_central_series (k : ℕ) : lie_submodule R L M :=
( ⊤ : lie_submodule R L M).lcs k
\end{lstlisting}
See also the final paragraph of section 3 in \cite{Nash2022}.

The lower central series is important because it provides a convenient way to define
nilpotency\extlink{algebra/lie/nilpotent.lean\#L172}:
\begin{lstlisting}
class is_nilpotent : Prop :=
(nilpotent : ∃ k, lower_central_series R L M k = ⊥)
\end{lstlisting}

Unwinding the definitions, informally a Lie module $M$ is nilpotent if and only if there exists
a natural number $k$ such that for any $x_1, x_2, \ldots, x_k$ in $L$ and any $m$ in $M$:
\begin{align}\label{nilpotentModuleEqn}
  [x_1, [x_2, \ldots, [x_k, m] \cdots ]] = 0.
\end{align}

We shall need the following property of a nilpotent Lie module: if $M$ is
non-trivial then it contains a non-zero element $m_0$ such that:
\begin{align}\label{nilpotentNonzeroKilled}
  [x, m_0] = 0 \quad\mbox{for all $x$ in $L$.}
\end{align}
Indeed one chooses a minimal $k$ satisfying \eqref{nilpotentModuleEqn} and
takes any non-zero element of the form:
\begin{align*}
  m_0 = [x_2, [x_3, \ldots, [x_k, m] \cdots ]].
\end{align*}
The result appears in Mathlib as\extlink{algebra/lie/nilpotent.lean\#L296}:
\begin{lstlisting}
lemma nontrivial_max_triv_of_is_nilpotent
  [nontrivial M] [is_nilpotent R L M] :
  nontrivial $ max_triv_submodule R L M :=
\end{lstlisting}
together with\extlink{algebra/lie/abelian.lean\#L110}:
\begin{lstlisting}
lemma mem_max_triv_submodule (m₀ : M) :
  m₀ ∈ max_triv_submodule R L M ↔ ∀ (x : L), ⁅x, m₀⁆ = 0 :=
\end{lstlisting}

\subsection{Acting nilpotently}
An element $a$ of an associative\footnote{We note in passing that
the concept generalises to non-associative rings and that the relation $[x, x] = 0$
means that all elements $x$ of a Lie algebra are nilpotent (taking $k = 2$). This fact
is partly responsible for the elevated role that nilpotency plays in Lie theory.} ring
$A$ is said to be nilpotent if there exists a natural number $k$ such that:
\begin{align*}
  a^k = 0.
\end{align*}

Given a Lie module $M$ of a Lie algebra $L$, this concept of nilpotency for elements of the
ring $A = \End(M)$ plays a central role in Engel's theorem:
\begin{definition*}
  We say a Lie algebra $L$ acts nilpotently on a Lie module $M$ if the image of the natural map:
  \begin{align*}
    \phi : L &\to \End(M)\\
    x &\mapsto \phi_x,
  \end{align*}
  contains only nilpotent elements.
\end{definition*}

A trivial but important observation is that if $M$ is a nilpotent Lie module then $L$
acts nilpotently. Indeed, given any $x$ in $L$, taking the constant sequence:
\begin{align*}
    x_1 = x_2 = \cdots = x_k = x,
  \end{align*}    
equation \eqref{nilpotentModuleEqn} reads $\phi_x^k(m) = 0$ for any $m$, and thus:
\begin{align*}
  \phi_x^k = 0.
\end{align*}
In Mathlib this statement is written\extlink{algebra/lie/nilpotent.lean\#L184}:
\begin{lstlisting}
lemma nilpotent_endo_of_nilpotent_module [is_nilpotent R L M] :
  ∃ (k : ℕ), ∀ (x : L), (to_endomorphism R L M x)^k = 0 :=
\end{lstlisting}
A corollary\footnote{Engel's theorem is stronger because it applies for any
$x_1, x_2, \ldots, x_k$, not just the constant sequence: $x_1 = x_2 = \cdots = x_k = x$.}
of Engel's theorem states that when $L$ is Noetherian as an $R$-module, the converse statement
holds. More precisely:
\begin{align*}
  \forall x, \exists k, \phi_x^k = 0 \iff \exists k, \forall x, \phi_x^k = 0.
\end{align*}

\section{Engel's theorem}\label{engelSect}
Our goal is to outline a proof of:
\begin{theorem*}[Engel's theorem]
  Let $M$ be a Lie module over a Lie algebra $L$ with coefficients in a commutative ring $R$.
  Suppose that $L$ is Noetherian as an $R$-module. Then $M$ is nilpotent if and only if
  $L$ acts nilpotently on $M$.
\end{theorem*}
The statement in Mathlib is\extlink{algebra/lie/engel.lean\#L274}:
\begin{lstlisting}
lemma is_nilpotent_iff_forall [is_noetherian R L] :
  lie_module.is_nilpotent R L M ↔
  ∀ x, is_nilpotent $ to_endomorphism R L M x :=
\end{lstlisting}

This should be compared with traditional versions appearing e.g. in
Serre \cite{serre1965} (I.4, Theorems 1, 2, 2'),
Bourbaki \cite{bourbaki1975} (I, 4.2, Theorem 1, Corollaries 1, 3),
Fulton, Harris, \cite{MR1153249} (9.2, Theorem 9.9, Exercise 9.10).
For the sake of definiteness we give such a statement.
\begin{theorem*}[Engel's theorem, traditional version]
  \leavevmode\newline
  \renewcommand{\theenumi}{(\alph{enumi})}
  \begin{enumerate}
    \item Let $L$ be a finite-dimensional Lie algebra with coefficients in a field.
    Then $L$ is nilpotent if and only if it acts nilpotently on itself.
    \item Let $M$ be a non-trivial finite-dimensional vector space
    and suppose that $L \subseteq \End(M)$ is a Lie subalgebra consisting of nilpotent elements.
    Then there exists a non-zero element $m_0$ in $M$ such that:
    \begin{align*}
      [x, m_0] = 0 \quad\mbox{for all $x$ in $L$.}
    \end{align*}
  \end{enumerate}
\end{theorem*}

Taking $M = L$ in our version recovers part (a) in the traditional statement. As explained
in the discussion of equation \eqref{nilpotentNonzeroKilled}, part (b) is also a trivial
corollary of our version. We have thus unified parts (a) and (b) in the traditional version
into a single result. Demanding the stronger property that $M$ is nilpotent also turns out to
simplify the proof as we get a stronger inductive hypothesis at the key step in the proof.

\subsection{Engelian Lie algebras}
It is convenient to introduce some terminology.
\begin{definition*}
  We say a Lie algebra $L$ is Engelian if any Lie module $M$ on which $L$ acts
  nilpotently is nilpotent.
\end{definition*}

The definition in Mathlib is\footnote{The universe \texttt{u₄} appearing is not significant and
merely reflects Mathlib's standard universe polymorphism. Lean can handle such polymorphism
automatically so one could (and probably should) replace \texttt{Type u₄} with
\texttt{Type*}.}\extlink{algebra/lie/engel.lean\#L158}:
\begin{lstlisting}
def lie_algebra.is_engelian : Prop :=
  ∀ (M : Type u₄) [add_comm_group M], by exactI ∀
    [module R M] [lie_ring_module L M], by exactI ∀
    [lie_module R L M], by exactI ∀
    (h : ∀ x, is_nilpotent $ to_endomorphism R L M x),
    lie_module.is_nilpotent R L M
\end{lstlisting}
In fact, this exposes a rough edge of Lean 3: quantification over typeclasses
requires use of the {\lq}\texttt{by exactI ∀} pattern{\rq} as well as the explicit mention of
typeclasses that could be inferred (like \texttt{add\_comm\_group}, \texttt{module} etc). One would prefer to write
simply \texttt{∀ (M : Type u₄), [[lie\_module R L M]]} rather than the
three lines appearing above. This situation is likely to improve after
Mathlib is migrated to Lean 4.

Using this new language, the non-trivial part of Engel's theorem is the statement that Noetherian
Lie algebras are Engelian:\extlink{algebra/lie/engel.lean\#L228}
\begin{lstlisting}
lemma is_engelian_of_is_noetherian [is_noetherian R L] :
  lie_algebra.is_engelian R L :=
\end{lstlisting}

\subsection{Passing to the image in End(M)}
We now fix a Noetherian Lie algebra $L$ together
with a Lie module $M$ on which it acts nilpotently. The first step in the proof is to
note that we may replace $L$ with its image:
\begin{align*}
  L' = \range \phi \subseteq \End(M),
\end{align*}
under the natural map:
\begin{align*}
  \phi : L \to \End(M).
\end{align*}
The relevant formal statement is\extlink{algebra/lie/nilpotent.lean\#L320}:
\begin{lstlisting}
lemma is_nilpotent_range_to_endomorphism_iff :
  is_nilpotent R (to_endomorphism R L M).range M ↔
  is_nilpotent R L M :=
\end{lstlisting}
The proof is essentially tautological but the observation is important and the
lemma is a good example of Lean's typeclass mechanism working well.

Without the
user doing anything, Lean recognises that \texttt{(to\_endomorphism R L M).range},
i.e. $L'$, is a Lie algebra and that $M$ is naturally a Lie module over $L'$. All
of this is achieved by typeclass instances registered far away, with no direct
connection to Engel's theorem.

The passage from $L$ to $L'$ is useful because of the following
lemma\extlink{algebra/lie/nilpotent.lean\#L693}:
\begin{lstlisting}
lemma lie_algebra.is_nilpotent_ad_of_is_nilpotent
  {A : Type v} [comm_ring R] [ring A] [algebra R A]
  {L' : lie_subalgebra R A} {x : L'}
  (h : is_nilpotent (x : A)) :
  is_nilpotent $ lie_algebra.ad R L' x :=
\end{lstlisting}
(Note that an associative algebra $A$ is naturally also a Lie
algebra\extlink{algebra/lie/of_associative.lean\#L105} which is
why we may write \texttt{lie\_subalgebra R A}. See also \cite{Nash2022} section 5.)
The proof follows from the binomial theorem.
Taking $A = \End(M)$ the lemma states that $L'$ acts nilpotently on itself.

\subsection{The main argument}
We introduce the set:
\begin{align*}
  s = \{K \subseteq L' ~|~ K \mbox{ is an Engelian Lie subalgebra} \},
\end{align*}
written in Mathlib as\extlink{algebra/lie/engel.lean\#L237}:
\begin{lstlisting}
  let s := { K : lie_subalgebra R L' | is_engelian R K },
\end{lstlisting}
Our goal is to show that $s$ contains $L'$, regarded as a Lie subalgebra of
itself. This is the top element $\top$ in the complete lattice of Lie subalgebras.

Since $s$ is non-empty (it contains the
trivial Lie subalgebra) and $L'$ is Noetherian, $s$ contains a maximal element.
We thus need only show that an element of $s$ cannot be maximal when it is a
proper subalgebra. In other words we must justify the following claim appearing
in the Mathlib proof\extlink{algebra/lie/engel.lean\#L243}:
\begin{lstlisting}
  have : ∀ (K ∈ s), K ≠ ⊤ → ∃ (K' ∈ s), K < K',
\end{lstlisting}

Thus let $K$ be a proper Engelian Lie subalgebra and consider the Lie
module $L' / K$ over $K$. Because $L'$ acts nilpotently on itself,
$K \subseteq L'$ acts nilpotently on $L'$ and thus also on $L' / K$.
Since $K$ is Engelian, $L' / K$ is nilpotent and thus
by \eqref{nilpotentNonzeroKilled} there exists $x \in L' - K$ such that:
\begin{align}\label{KxEqn}
  [K, x] \subseteq K.
\end{align}

Finally let:
\begin{align*}
  K' = \{ \lambda x + \mu k ~|~ k \in K; \lambda, \mu \in R \}.
\end{align*}
Using \eqref{KxEqn} and recalling that $[x, x] = 0$ we see that:
\par\vskip\baselineskip
\begin{itemize}
  \itemsep1em 
  \item $K'$ is a Lie subalgebra of $L'$,
  \item $K'$ strictly contains $K$,
  \item $K$ is an ideal in $K'$.
\end{itemize}
\par\vskip\baselineskip

To finish the argument we need only show that $K'$ is Engelian.
We thus fix a Lie module $M'$ over $K'$ on which $K'$ acts nilpotently. We must show that $M'$
is nilpotent. To do so, we apply the following lemma
(whose proof is a simple induction)\extlink{algebra/lie/engel.lean\#L107}:
\begin{lstlisting}
lemma lcs_le_lcs_of_is_nilpotent_span_sup_eq_top
  {I : lie_ideal R L} {x : L}
  (hxI : (R ∙ x) ⊔ I = ⊺)
  {n i j : ℕ}
  (hxn : (to_endomorphism R L M' x)^n = 0)
  (hIM : lower_central_series R L M' i ≤ I.lcs M' j) :
  lower_central_series R L M' (i + n) ≤ I.lcs M' (j + 1) :=
\end{lstlisting}
with the roles of $I$, $L$ played by $K$, $K'$.

We see that if the lower central series of $M'$ as a Lie module over $K$ reaches
zero, then its lower central central series over $K'$ must also reach zero. Thus
nilpotency over $K$ (which follows since $K$ is Engelian) ensures nilpotency over $K'$,
as required.

\section{The zero root space}\label{rootSpaceSect}
As noted in \cite{Nash2022} section 11, Mathlib contains definitions of and basic results about
weight spaces and root spaces. Engel's theorem was added to Mathlib for the purpose
of advancing this theory. It may be instructive to outline this application.

Given a nilpotent Lie subgalgebra $H \subseteq L$,
a distinguished role is played by the zero root space. In Mathlib this is
\texttt{root\_space H (0 : H → R)}\extlink{algebra/lie/weights.lean\#L276}; informally it is:
\begin{align*}
  L^0 = \{ x ~|~ \exists k, \forall y \in H, \ad_y^k(x) = 0 \} \subseteq L.
\end{align*}
Any root space is a $H$-submodule of $L$ but more is true for the zero root space;
it is a Lie subalgebra\extlink{algebra/lie/weights.lean\#L383}:
\begin{lstlisting}
def zero_root_subalgebra : lie_subalgebra R L :=
{ lie_mem' := ...,
  .. (root_space H 0 : submodule R L) }
\end{lstlisting}
One always has $H \subseteq L^0$ and it is easy to see that if $H = L^0$ then
$H$ is a Cartan subalgebra. When $L$ is Noetherian, Engel's theorem shows that
the converse holds:
\begin{align}\label{zeroRootCartanIff}
  H = L^0 \iff \mbox{$H$ is a Cartan subalgebra.}
\end{align}
The literature appears only to contain proofs of this result when the coefficients
are a field. Moreover the arguments in the literature cannot be used since they
argue inductively using dimension --- a concept that does not exist over
general coefficients. Fortunately with our version of Engel's theorem, the proof is
almost trivial.

Here is the statement of \eqref{zeroRootCartanIff} in Mathlib\extlink{algebra/lie/weights.lean\#L468}:
\begin{lstlisting}
lemma zero_root_subalgebra_eq_iff_is_cartan [is_noetherian R L]:
  zero_root_subalgebra R L H = H ↔ H.is_cartan_subalgebra :=
\end{lstlisting}
The point is that if $L$ is Noetherian, then $H$ acts nilpotently on $L^0$ and so
by Engel's theorem, $L^0$ is nilpotent as a Lie module over $H$. Mathlib actually
contains the following slightly more general result which holds for any
Noetherian Lie module $M$\extlink{algebra/lie/weights.lean\#L262}:
\begin{lstlisting}
instance [lie_algebra.is_nilpotent R L]
  [is_noetherian R L] [is_noetherian R M] :
  is_nilpotent R L $ weight_space M (0 : L → R) :=
\end{lstlisting}
Once we know that $L^0$ is nilpotent over $H$, the self-normalizing property of
a Cartan subalgebra ensures that $H$ cannot be a strict submodule and the result
is immediate.

\section{Conclusion}
Having extolled the virtues of generality and unity, we should emphasise that
in some sense they are two sides of the same coin. For example, the theory of
root spaces discussed here depends on Mathlib's theory of eigenspaces.
This same theory of eigenspaces is also used in Mathlib's library of functional
analysis \cite{DupuisLewisMacbeth2022}. This is only possible because
the theory of eigenspaces was developed quite generally (before either application
existed). We thus enjoy a synergy: any lemma added by one consumer of the eigenspace
theory is automatically available to the other.

Although we have stated the required finiteness condition for Engel's theorem as the
requirement that $L$ be Noetherian as an $R$ module, the argument actually needs
only the weaker condition that the lattice of Lie subalgebras of $L$
possesses a maximality property\footnote{I.e., any non-empty subset of Lie subalgebras
contains a maximal element.}.

Our version of Engel's theorem seems not to exist in the literature.
One notable close pass is an old paper of Zorn \cite{Zorn1937} where
the result is proved for Lie rings (but not for their Lie modules). Another
interesting paper is Jacobson \cite{Jacobson1952}. In place of our:
\begin{align*}
  L' \subseteq \End(M),
\end{align*}
he proves Engel's theorem for what he calls a $\Phi$-subring:
\begin{align*}
  \mathfrak{A} \subseteq \mathfrak{U}.
\end{align*}
On the one hand Jacobson's result is more general since the class of $\Phi$-subrings
includes Lie subrings of associative rings as a special case. On the other hand
it is less general since he obtains his result subject to the much more restrictive
assumption that the lattice of ideals of the enclosing ring $\mathfrak{U}$ possesses
a minimality condition whereas we require only that the lattice of subalgebras of $L'$
possesses a maximality condition.

Finally we should mention that there is an important setting where Engel-type
results for Lie rings play a central role. See for example Zel'manov's
breakthrough \cite{Zelmanov1990}.

\vskip\baselineskip
\textbf{Acknowledgements}
\begin{quotation}
  \emph{I am grateful to Kevin Buzzard at Imperial College whose Excellence Fund in
  Frontier Research grant funded this work. I am also grateful to the anonymous referees
  for numerous helpful remarks.
  I declare that this article is the sole work of the named author
  and that I have no conflicts of interest pertaining to any of the work
  discussed here. All code is available in the \texttt{master} branch of the open-source
  \texttt{Mathlib} repository, with links appearing throughout the text.}
\end{quotation}

\bibliography{paper}

\end{document}